\documentclass{ifacconf}

\usepackage{cite}
\usepackage{natbib}

\usepackage{amssymb,amsmath}
\usepackage{dsfont,mathrsfs}
\usepackage{amsbsy}

\usepackage{graphicx}
\usepackage{graphics} 
\usepackage{epsfig}
\usepackage{subfigure}
\usepackage{adjustbox}
\graphicspath{{images/}}
\usepackage{empheq}
\usepackage{color}
\usepackage{xcolor}

\usepackage{comment}

\DeclareMathAlphabet{\mathbfcal}{OMS}{cmsy}{b}{n}
\DeclareFontFamily{T1}{calligra}{}
\DeclareFontShape{T1}{calligra}{m}{sl}{<->s*[1]callig15}{}  %
\DeclareMathAlphabet\mathcalligra{T1}{calligra}{m}{sl}

\newtheorem{Prb}{{Problem}}
\newtheorem{Lem}{{Lemma}}
\newtheorem{Rem}{\textbf{Remark}}

\newcommand{\vect}[1]{\mathbf{#1}}
\newcommand{\mat}[1]{\mathbf{#1}}

\renewcommand{\imath}{\boldsymbol{\mathrm{i}}}

\newcommand{\colvec}[2][.9]{%
	\scalebox{#1}{%
		\renewcommand{\arraystretch}{.9}%
		$\begin{bmatrix}#2\end{bmatrix}$%
	}
}

\usepackage{colortbl} 

\usepackage{xparse}
\NewDocumentCommand{\ceil}{s O{} m}{%
	\IfBooleanTF{#1} 
	{\left\lceil#3\right\rceil} 
	{#2\lceil#3#2\rceil} 
}
\NewDocumentCommand{\floor}{s O{} m}{%
	\IfBooleanTF{#1} 
	{\left\lfloor#3\right\rfloor} 
	{#2\lfloor#3#2\rfloor} 
}

\renewcommand{\deg}[1]	{\mathrm{deg}({#1})} 
\def \dgr 				{\mathbf{D}} 
\def \normlap		 	{\boldsymbol{\mathcal{L}}} 

\def\cv{{\bf c}}

\begin{document}
\begin{frontmatter}

\title{A Proximal Point Approach for \\ Distributed System
State Estimation}

\thanks[footnoteinfo]{Part of this work was supported by MIUR (Italian Ministry for Education) under the initiative \textquotedblleft Departments of Excellence".}

\author[First]{Marco Fabris} 
\author[First]{Giulia Michieletto} 
\author[First]{Angelo Cenedese} 

\address[First]{Dept. of Information Engineering, University of Padova, Italy \\ (corresponding author e-mail: marco.fabris.7@phd.unipd.it)}

%
%
%

\begin{abstract}          
System state estimation constitutes a key problem in several applications involving multi-agent system architectures. This rests upon the estimation of the state of each agent in the group, which is supposed to access only relative measurements w.r.t. some neighbors state. 
%
%
Exploiting the standard least-squares paradigm, the system state estimation task is faced in this work by deriving 
a distributed Proximal Point-based iterative scheme. 
This solution entails the emergence of interesting connections between the structural properties of the stochastic matrices describing the system dynamics and the convergence behavior toward the optimal estimate. A deep analysis of such relations is provided, jointly with a further discussion on the penalty parameter that characterizes the Proximal Point approach.
\end{abstract}


\end{frontmatter}

\section{Introduction}

In the last twenty years, the advancements in pervasive computing and ambient intelligence have led to a fast development of smart networks characterized by the presence of various interacting cyber-physical systems (as, e.g., smart grids and robotic networks)  
whose effectiveness rests  upon the cooperation of multiple autonomous entities under constraints of real-time high-level performance. In this perspective, the role of multi-agent systems is evident, due to the capability of these architectures involving large sets of interactive devices to solve complex tasks by exploiting local computation and communication~\citep{chen2019control}.


One of the main challenging aspects of the embedded multi-agent systems consists in the wise management and exploitation of the local information since they are typically involved in the solution of large-scale logically/spatially distributed optimization problems. Along this line, \textit{system state estimation} constitutes a popular task in several control and learning smart networks applications, requiring to estimate the state of each agent through the adoption of a distributed paradigm exploiting local information and interactions~(see, e.g., \citep{cintuglu2019secure} 
wherein the task is faced accounting for network microgrids). 

Many existing works explore the properties of the distributed state estimation solution w.r.t. the topological features of the graph modeling the network in terms of available measurements. For instance, 
in~\citep{barooah2009error}, the variance of the estimation error for a node variable is shown to increase proportionally to the distance of the node itself w.r.t an arbitrary reference one, hence a topological classification is provided resting upon the graph sparsity. In \citep{carron2014asynchronous}, instead, the system state problem boils downs to the agent position estimation exploiting some relative noisy distances measurement. This task is tackled through a consensus-based algorithm whose rate of convergence in expectation of the estimation error is determined for regular  graphs, accounting for the network size. 
Recently, in~\citep{ravazzi2018distributed}, the problem of distributed estimation is studied accounting for relative measurements characterized by heterogeneity and uncertainty. In this case a Least-Square (LS) approach is adopted. Similarly, in~\citep{fabris2019distributed}, the problem is solved proposing a decentralize iterative solution whose convergence features are strictly related to the spectral properties of the update matrix.

The major contribution of this work consists in the design of a distributed iterative procedure for system state estimation. Its novelty relies on the use of Proximal Point (PP) method, 
 nowadays widely employed in different research areas (such as statistical learning
 \citep{bai2019general})
 and recently exploited in  the development of algorithms for the solution of estimation tasks resting upon the LS approach~\citep{aster2018parameter}. 
A deep insight is provided on the role of topological links that define the communication constraints in the system w.r.t. the convergence properties of the proposed method, along with a convergence analysis that guarantees the scalability of the proposed solution. 
Moreover, a discussion on the role of the penalty parameter -- as usual in many related works, see also \citep{olfati2007consensus} -- characterizing the PP approach is proposed.

The rest of the paper is organized as follows. In Sec.~\ref{sec:problem_statement} the system state estimation is formally stated as a minimization problem. 
In Sec. \ref{sec:problem_solution} its distributed PP-based solution is derived, while Sec. \ref{sec:convergence_analysis} is devoted to the convergence analysis of the proposed solution. The theoretical observations are then validated via numerical simulations for different network topologies in Sec.~\ref{sec:numerical_simulation}. Finally, in Sec. \ref{sec:conclusions} conclusions are drawn and future work are envisaged.

\section{System State Estimation Problem}
\label{sec:problem_statement}

This section is devoted to the formalization of the system state estimation task as an optimization problem based on the set of the available relative measurements. In doing this, a graph-based representation is introduced for multi-agent systems that model the devices interactions.

\subsection{Graph-Based Network Model}
\label{sec:graph_model}

An $n$-agent system can be modeled through a graph $\mathcal{G}=\left(\mathcal{V},\:\mathcal{E}\right)$ so that each element in the vertex set $\mathcal{V}=\left\{v_1, \ldots, v_n\right\}$ is related to an agent in the group, while the edge set $\mathcal{E}\subseteq \mathcal{V}\times \mathcal{V}$ characterizes the agents interactions in terms of both sensing and communication capabilities. Hereafter, bidirectional agents interactions are supposed, therefore $\mathcal{G}$ is assumed to be undirected and such that $e_{ij}=\{v_i , v_j\} \in \mathcal{E}$ if and only if the $i$-th and $j$-th agents can sense each other and are able to mutually exchange information according to some communication protocol.

Agent links are often described by the \textit{adjacency matrix} $\mathbf{A} \in \mathbb{R}^{n \times n}$ such that $[\mathbf{A}]_{ij}=1$ if $e_{ij}\in\mathcal{E}$ and $[\mathbf{A}]_{ij}=0$ otherwise. The set $\mathcal{N}_i=\left\{v_j\in\mathcal{V}, j \neq i \;|\; [\mathbf{A}]_{ij}=1\right\}$ thus identifies the \textit{neighborhood} of the node $v_i$, i.e., the set of agents interacting with the $i$-th one. The cardinality $\mathrm{deg}(v_i)$ of $\mathcal{N}_i$ is generally referred to as \textit{degree} of the $i$-th agent, and $d_{m}={\min}_{i \in \{1,\ldots, n\}}\{\text{deg}(v_i)\}$ and $d_{M}={\max}_{i \in \{1,\ldots, n\}}\{\text{deg}(v_i)\}$ respectively identify the \textit{minimum} and \textit{maximum} degree of the graph. Furthermore, the degree of the $i$-th agent corresponds to the $i$-th element of the  (diagonal) \textit{degree matrix} $\mathbf{D} = \text{diag}(\mathbf{A}\mathds{1}_n)\in \mathbb{R}^{n\times n}$, where $\mathds{1}_n \in \mathbb{R}^n$ indicates a $n$-dimensional (column) vector whose entries are all ones. In particular, matrix $\mathbf{D}$ contributes to the definition of the \textit{Laplacian matrix}, $\mathbf{L} = \mathbf{D}-\mathbf{A} \in \mathbb{R}^{n \times n}$, that summarizes the topology of the networked system. 


\subsection{Problem Statement}
\label{sec:theOptimizationProblem}

According to the introduced graph-based model, let us consider a multi-agent system composed of $n$ devices whose interactions are represented by an undirected \textit{connected} graph $\mathcal{G}$. Each $i$-th agent, $i \in \{ 1, ..., n \}$, in the network is assumed to be characterized by a scalar attribute $x_i \in \mathbb{R}$ (the \textit{$i$-th agent state}) corresponding to a physical quantity, and by a set of relative measurements of the same quantity, namely $\{\tilde{x}_{ij} \in \mathbb{R}, \; v_j \in \mathcal{N}_i\}$ where $\tilde{x}_{ij}$ represents a noisy measurement of the difference between $x_j$ and $x_i$. 

Given these premises, the system state estimation problem   consists in the determination of the agents state set $\{x_1^\ast, \ldots, x_n^\ast \}$ that allows to best approximate (and be consistent with) the set of the existing measurements. 
Adopting the standard LS approach, this estimation issue can be formalized in the  optimization framework

\begin{Prb}
\label{pr:system_state_estimation}
For a given $n$-agent networked system modeled by the graph $\mathcal{G}$, let  $\cup_{i = 1, \ldots, n} \{\tilde{{x}}_{ij} \in \mathbb{R}, \; v_j \in \mathcal{N}_i\}$ be the set of all the available relative measurements. Then, the system state estimation task consists in the resolution of the following convex minimization problem
	\begin{align}
	\label{eq:cost}
	\underset{\{x_1, \ldots, x_n\}}{\arg \min} & \; \frac{1}{2} \textstyle\sum\limits_{v_i \in \mathcal{V}} \textstyle\sum\limits_{v_j \in \mathcal{N}_i} ( x_i-x_j+\tilde{x}_{ij} )^2,
	\end{align}
whose cost function is hereafter denoted by $h(\mathbf{x})$, with  $\mathbf{x} = \colvec{ x_1 \;\; \cdots \;\;  x_n}^\top \in \mathbb{R}^n$  identifying the \textit{system state vector}.
\end{Prb} 



\section{Distributed PP-Based Solution}
\label{sec:problem_solution}

In this section, the main contribution of this work is provided presenting a distributed PP-based solution for Problem~\ref{pr:system_state_estimation} inspired by the state-of-art centralized and distributed approaches described in~\citep{fabris2019distributed}.

\subsection{Preliminary Observations}
\label{sec:centralized_distributed_solution}

 Problem~\ref{pr:system_state_estimation} can be solved in a centralized way through the computation of the gradient of the cost function $h(\vect{x})$ w.r.t. $\mathbf{x}$, whose $i$-th component, $i \in \{ 1, \ldots, n\}$ results in
\begin{equation*}
[\nabla_{\mathbf{x}}h(\vect{x})]_i = 
2\mathrm{deg}(v_i)x_i - 2 \!\!\textstyle\sum\limits_{v_j \in {\mathcal N}_i}\!\! x_j - \!\!\textstyle\sum\limits_{v_j \in {\mathcal N}_i}\!\!(\tilde{x}_{ji} - \tilde{x}_{ij}).
\label{eq:costGraSO2}
\end{equation*}
In particular, imposing the equilibrium condition $\nabla_{\mathbf{x}}h(\vect{x})  = \mathbf{0}_{n}$, with $\mathbf{0}_{n} \in \mathbb{R}^n$ indicating the $n$-dimensional zero (column) vector, the following linear system is derived
\begin{align}
\label{eq:systemBetaDeltaBeta}
2 \mathbf {L}\mathbf{x} = \tilde{\mathbf{x}},
\end{align}
where $\tilde{\mathbf{x}}  = \colvec{ \textstyle\sum\nolimits_{v_j \in {\cal N}_1}(\tilde{x}_{j1} -\tilde{x}_{1j}) \;\; \cdots \;\; \textstyle\sum\nolimits_{v_j \in {\cal N}_n}(\tilde{x}_{jn} - \tilde{x}_{nj})}^\top \in\mathbb{R}^n$ is the \textit{relative measurements vector} and $\mathbf {L} \in \mathbb{R}^{n \times n}$ is the Laplacian matrix associated to $\mathcal{G}$. 	
In the light of~\eqref{eq:systemBetaDeltaBeta}, it is straightforward that, if $\tilde{\mathbf{x}} \notin \ker(\mathbf{L}) \setminus \left\lbrace \mathbf{0}_{n} \right\rbrace$, the (minimum norm) \textit{centralized} solution $\mathbf{x}^\star \in \mathbb{R}^n$ of the minimization~\eqref{eq:cost} can be computed accounting for the pseudo-inverse $\mathbf{L}^{\dagger}  \in \mathbb{R}^{n \times n}$ of the Laplacian matrix, namely
\begin{equation}
\mathbf{x}^\star= \frac{1}{2} \mathbf{L}^{\dagger} {\tilde{\mathbf{x}}}. 
\label{eq:centrSol}
\end{equation}

On the other hand, Problem~\ref{pr:system_state_estimation} can be solved also adopting the distributed paradigm. Indeed,
 each equation of system~\eqref{eq:systemBetaDeltaBeta} is related to  single agent information in terms of local topology and measurements. In particular, by setting $[\nabla_{\mathbf{x}} h(\vect{x})]_i = 0$, $i \in \{1, \ldots, n\}$, it follows that
\begin{equation}
x_i =  \frac{2\sum_{v_j \in {\cal N}_i} x_j + \sum_{v_j \in {\cal N}_i}(\tilde{x}_{ji} - \tilde{x}_{ij})}{2\mathrm{deg}(v_i)},
\label{eq:distrUpdateScalar}
\end{equation}
namely, at the equilibrium, the $i$-th agent state depends only on the state  of its neighbors and on its set of relative measurements. 
Thus, to solve Problem~\ref{pr:system_state_estimation}, the following  iterative update rule can be defined for the whole system: 
\begin{equation}
\Sigma_{0}: \quad \mathbf{x}(k+1) = \mathbf{F}_{0} \mathbf{x}(k) + \mathbf{u}_{0},
\label{eq:distrUpdateVector1}
\end{equation}
where the state matrix $\mathbf{F}_{0} \in \mathbb{R}^{n \times n}$ is derived by (pre)-normalizing the adjacency matrix through the degree matrix, i.e.,
\begin{equation*}
\label{eq:matrix_F}
\mathbf{F}_{0}=\mathbf{D}^{-1} \mathbf{A}.
\end{equation*}
Similarly, the input $\mathbf{u}_{0} \in \mathbb{R}^n$ in~\eqref{eq:distrUpdateVector1} is determined by the normalizing relative measurements vector, i.e.,
\begin{equation*}
\mathbf{u}_{0} = \dfrac{1}{2}\mathbf{D}^{-1}\tilde{\mathbf{x}}.
\label{eq:inputvector}
\end{equation*}

Note that, since the graph $\mathcal{G}$ describing the multi-agent system is assumed to be connected, the spectrum $\Lambda(\mathbf{F}_{0})$ of $\mathbf{F}_{0}$ accounts for $n$ real eigenvalues $\lambda^{\mathbf{F}_{0}}_0 \geq \cdots \geq \lambda^{\mathbf{F}_{0}}_{n-1}$ in the range $\left[-1,1\right]$, with $\lambda^{\mathbf{F}_{0}}_0=1$ having single algebraic multiplicity. 
In addition, in~\citep{fabris2019distributed}, it is shown that the convergence of system~\eqref{eq:distrUpdateVector1} is ensured only if $\lambda_{n-1}^{\mathbf{F}_{0}} \neq -1$. This condition is fulfilled if and only if $\mathcal{G}$ is not bipartite, because the spectrum  of the \textit{normalized Laplacian matrix} $\normlap=\mathbf{I}_n-\dgr^{-1/2} \mathbf{A} \dgr^{-1/2} \in \mathbb{R}^{n \times n}$, corresponding to the eigenvalues  $0=\lambda_{0}^{\mathbfcal{L}} < \lambda_{1}^{\mathbfcal{L}} \leq \dots \leq \lambda_{n-1}^{\mathbfcal{L}} \leq 2$ is related to $\Lambda(\mathbf{F}_{0})$ so that
	\begin{equation}
	\label{eq:correspondance_between_lapl_and_F_eigs_}
	\lambda_{i}^{\mathbfcal{L}}=1-\lambda_{i}^{\mathbf{F}_{0}}, \quad i \in \{0, \ldots, n-1\}.
	\end{equation}

\subsection{Solution Derivation}\label{subsec:alg_rho}
%
%
%
%

In the following, an alternative iterative distributed solution is proposed for Problem~\ref{pr:system_state_estimation}. This rests on the PP approach illustrated  in~\citep{parikh2014proximal} and entails the presence of the penalty parameter $\rho \geq 0$ whose tuning, depending on the network topology, may improve the convergence properties toward the optimal solution as compared to~\eqref{eq:distrUpdateScalar}, as will be clarified in the next section.

Let us consider then the \textit{regularized PP objective function} $h_\rho(k)=h_{\rho}(\mathbf{x}(k),\mathbf{x}(k+1))$ depending on the system state at the $k$-th and $(k+1)$-th step as follows
\begin{equation*} \label{psi_k}
h_{\rho}(k) = h(\mathbf{x}(k+1)) + \dfrac{\rho}{2} \left\| \mathbf{x}(k+1)-\mathbf{x}(k) \right\|_{2}^{2},
\end{equation*}
with $\rho \geq 0$.
The gradient and Hessian of such a function w.r.t. $\mathbf{x}(k+1)$ result to be respectively
\begin{align}
&\nabla_{\mathbf{x}(k+1)} h_{\rho}(k) = \nabla_{\mathbf{x}(k+1)} h(\vect{x}(k+1)) + \rho (\mathbf{x}(k+1)-\mathbf{x}(k)), \nonumber\\
&\mathcal{H}_{\mathbf{x}(k+1)\mathbf{x}(k+1)} h_{\rho}(k) = 2\mathbf{L} + \rho \boldsymbol{I}_{n}. \label{ineq:Hxxhrho}
\end{align} 
Note that, for $\rho > 0$, the Hessian~\eqref{ineq:Hxxhrho} is positive definite\footnote{Let $\mathbf{M}\in \mathbb{R}^{p \times p}$ be a matrix. Notation $\mathbf{M} \succeq 0$ and $\mathbf{M} \succ 0$ implies that $\mathbf{M}$ is positive semidefinite and positive definite in a generalized sense, respectively, i.e., all eigenvalues of $\mathbf{M}\succeq 0$ are real and nonnegative, even when $\mathbf{M}\neq \mathbf{M}^{\top}$ (bilinear forms in the latter case are not addressed). Moreover, given two $p \times p$ matrices $\mathbf{M}_{1}$ and  $\mathbf{M}_{2}$, the notation $\mathbf{M}_{1} \succeq \mathbf{M}_{2}$ stands for $\mathbf{M}_{1}-\mathbf{M}_{2} \succeq 0$.} ($\mathcal{H}_{\mathbf{x}(k+1)\mathbf{x}(k+1)} h_{\rho}(k)\succ 0$), hence 
the cost function $h_\rho (k)$ 
turns out to be strictly convex and its minimization admits a unique global solution. This can be derived imposing the equilibrium condition
$\nabla_{\mathbf{x}(k+1)} h_{\rho}(k)=\mathbf{0}_{n}$, which, for all $i\in \{ 1, \ldots, n\}$, yields
\begin{equation}
\label{eq:solution_min_ADMM_Boyd}
x_{i}(k+1) =  \dfrac{\rho x_{i}(k) + 2\sum_{v_j \in {\cal N}_i} x_j(k)   + \sum_{v_j \in {\cal N}_i}(\tilde{x}_{ji} - \tilde{x}_{ij})}{2\deg{v_{i}} + \rho}.
\end{equation}

Exploiting~\eqref{eq:solution_min_ADMM_Boyd}, the following update rule can be designed:
\begin{equation} \label{eq:distrUpdateVector_rho}
\Sigma_{\rho}: \quad \mathbf{x}(k+1) = \mathbf{F}_{\rho} \mathbf{x}(k) + \mathbf{u}_{\rho}.
\end{equation}
In this case, the state matrix $\mathbf{F}_{\rho} \in \mathbb{R}^{n \times n}$ is equal  to the $\rho$-regularized adjacency matrix (pre)-normalized by the $\rho$-regularized degree matrix, namely 
\begin{equation}\label{eq:Frho}
\mathbf{F}_{\rho} = \left( \mathbf{D} + \dfrac{\rho}{2} \mathbf{I}_{n}  \right)^{-1} \left(  \mathbf{A} + \dfrac{\rho}{2} \mathbf{I}_{n}    \right).
\end{equation}
Similarly, the input $\mathbf{u}_{\rho} \in \mathbb{R}^n$ in \eqref{eq:distrUpdateVector_rho} is given by the vector of the $\rho$-regularized and normalized relative measurements, i.e.,
\begin{equation*}\label{eq:input_rho}
\mathbf{u}_{\rho} =  \dfrac{1}{2}\left( \mathbf{D} + \dfrac{\rho}{2} \mathbf{I}_{n}  \right)^{-1} \tilde{\mathbf{x}} =  \left( \mathbf{D} + \dfrac{\rho}{2} \mathbf{I}_{n}  \right)^{-1} \mathbf{D} \mathbf{u}_{0}.
\end{equation*}

Accounting for~\eqref{eq:distrUpdateVector_rho}, the $i$-th agent state estimate at the $(k+1)$-th step turns out to be affected by the neighbors relative measurements and the neighbors state estimates at $k$-th step and by its relative measurements and the self estimate at $k$-th step. In particular, the self estimate is weighted through the penalty parameter $\rho$ that thus allows to define a certain level of confidence w.r.t. the previous step computation. Note that, for $\rho = 0$, the proposed scheme $\Sigma_{\rho}$ corresponds to $\Sigma_{0}$ in \eqref{eq:distrUpdateVector1}. 

To conclude, because the graph $\mathcal{G}$ representing the multi-agent system is supposed to be connected, the spectrum $\Lambda(\mat{F}_\rho)$ of the (row-stochastic) matrix $\mathbf{F}_{\rho}$ consists in $n$ real eigenvalues $\lambda^{\mathbf{F}_{\rho}}_0 \!\geq\! \cdots \!\geq\! \lambda^{\mathbf{F}_{\rho}}_{n-1}$ belonging to the set $((\rho - 2d_{M})/(\rho+2d_{M}) ,1] \!\subseteq\! (-1,1]$, for $\rho>0$, with $\lambda^{\mathbf{F}_{\rho}}_0=\lambda^{\mathbf{F}_{0}}_0=1$ having single algebraic multiplicity.

\section{Convergence Analysis}
\label{sec:convergence_analysis}


Adopting the update rule~\eqref{eq:distrUpdateVector_rho}, the convergence of the agent state estimate toward the equilibrium solution depends on the spectrum of the state matrix $\mathbf{F}_{\rho}$ and, in particular, the convergence rate depends on its second largest eigenvalue in modulus. Thus, this section aims at investigating the properties of $\Lambda(\mat{F}_\rho)$ and, subsequently, discussing the performance of the scheme $\Sigma_\rho$ based on the topology characterizing the multi-agent system and on the selection of the penalty parameter $\rho$.

\subsection{Insight on the Spectral Properties of $\mat{F}_\rho$} 
\label{State-space Model sigma_rho}

First, note that the spectrum of the matrix $\mathbf{F}_{\rho}$ in~\eqref{eq:Frho} trivially depends on the parameter $\rho$. In particular, property stated in the next lemma holds.

\begin{Lem}\label{lem:bounds_on_eigs_F_rho}
Considering two values $\rho_1, \rho_2$  of the parameter $\rho$, it holds that $\lambda_{i}^{\mathbf{F}_{\rho_{1}}}   > \lambda_{i}^{\mathbf{F}_{\rho_2}}$ for $i \in \{1, \ldots, n-1\}$ if and only if $\rho_1 >\rho_2 \geq 0$. 
\end{Lem}
\begin{pf}
Consider $\mathbf{F}_{\rho}$ as the continuous function $\mathbb{R}_{>0} \rightarrow \mathbb{R}^{n\times n}, \rho \mapsto \left( 2\mathbf{D} + \rho \mathbf{I}_{n}  \right)^{-1} \left(  2\mathbf{A} + \rho \mathbf{I}_{n}    \right)$. By the Gershgorin circle theorem, the   derivative of $\mathbf{F}_{\rho}$ w.r.t. $\rho$ is positive semidefinite, namely $\mathbf{F}^{\prime}_{\rho}  =  d\mathbf{F}_{\rho}/d \rho = 2\left( 2\mathbf{D} + \rho \mathbf{I}_{n}  \right)^{-2} \mathbf{L} \succeq 0.$
Hence, $\mathbf{F}_{\rho}$ is a nondecreasing function of the parameter $\rho$, i.e., $\rho_1 > \rho_{2} \geq 0 $ implies that $\mathbf{F}_{\rho_1} \succeq \mathbf{F}_{\rho_{2}} $ and viceversa. As a consequence, it follows that  $\lambda_{i}^{\mathbf{F}_{\rho_{1}}}   > \lambda_{i}^{\mathbf{F}_{\rho_2}}$ for $i \in \{1, \ldots, n-1\}$ if and only if $\rho_1 >\rho_2 \geq 0$.	\hfill {$\square$}
\end{pf}

Exploiting the results of Lem.~\ref{lem:bounds_on_eigs_F_rho}, the next proposition provides a lower and an upper bound for each eigenvalue of $\mathbf{F}_{\rho}$, which  turn out to be functions of the eigenvalues of $\mathbf{F}_{0}$, of the parameter $\rho$ and of the maximum $d_M$ and minimum $d_m$ degree of $\cal G$.

\begin{prop}\label{prop:bounds_on_eigs_F_rho}
For any $\rho > 0$, all the eigenvalues $\lambda^{\mathbf{F}_{\rho}}_{i}$, $i \in \{ 1, \ldots, n-1 \}$, of matrix $\mathbf{F}_{\rho}$ admit the following lower $\underline{\lambda}^{\mathbf{F}_{\rho}}_{i}$ and upper $\overline{\lambda}^{\mathbf{F}_{\rho}}_{i}$ bounds:
	\begin{equation}\label{eq:upper_lower_bound_lambda_i_F_rho}
	\underline{\lambda}^{\mathbf{F}_{\rho}}_{i} = \dfrac{\rho+2 \lambda^{\mathbf{F}_{0}}_{i} d_{M}}{\rho+2d_{M}} \quad \text{and} \quad \overline{\lambda}^{\mathbf{F}_{\rho}}_{i}=\dfrac{\rho+ 2\lambda^{\mathbf{F}_{0}}_{i} d_{m}}{\rho+2d_{m}}.
	\end{equation}
Trivially, it holds that $\underline{\lambda}^{\mathbf{F}_{\rho}}_{i} = \lambda^{\mathbf{F}_{\rho}}_{i} =\overline{\lambda}^{\mathbf{F}_{\rho}}_{i}$ if and only if the graph $\mathcal{G}$ is regular, i.e., $\mathrm{deg}(v_i)=d$ $ \forall v_i \in \mathcal{V}$ with $d>0$.
\end{prop}

\begin{pf}
	Let $\mathbf{\tilde{F}}_{\rho}(\mathbf{M}):\mathbb{R}^{n\times n}\rightarrow\mathbb{R}^{n\times n}$ be a function that maps a diagonal matrix $\mathbf{M}\succ 0$ to the matrix $\mathbf{\tilde{F}}_{\rho}(\mathbf{M})= \left( 2\mathbf{M}+\rho \mathbf{I}_{n}  \right)^{-1} \left( 2\mathbf{M} \mathbf{F}_{0}+\rho \mathbf{I}_{n}  \right).$
Exploiting~\eqref{eq:correspondance_between_lapl_and_F_eigs_}, criteria in \cite{drazin1962} and the Gershgorin circle theorem, it is possible to prove that the condition $d_{M} \mathbf{I}_{n} \succeq \mathbf{M} \succeq d_{m} \mathbf{I}_{n}$, i.e. $\mathbf{M} - d_{m} \mathbf{I}_{n} \succeq 0$ and $d_{M}\mathbf{I}_{n}-\mathbf{M} \succeq 0$, implies
	\begin{align}
&\begin{cases}
2\rho(\mathbf{M}-d_{m}\mathbf{I}_{n})(\mathbf{I}_{n}-\mathbf{F}_{0}) \succeq 0 \\
2\rho(d_{M}\mathbf{I}_{n}-\mathbf{M})(\mathbf{I}_{n}-\mathbf{F}_{0}) \succeq 0 
\end{cases}\nonumber\\
	& \begin{cases}
		(2\mathbf{M}+\rho \mathbf{I}_{n})(2d_{m}\mathbf{F}_{0}+\rho\mathbf{I}_{n}) \succeq (2d_{m} +\rho )(2\mathbf{M}\mathbf{F}_{0} + \rho \mathbf{I}_{n}) \\
		(2d_{M} +\rho )(2\mathbf{M}\mathbf{F}_{0} + \rho \mathbf{I}_{n}) \succeq   (2\mathbf{M}+\rho \mathbf{I}_{n})(2d_{M}\mathbf{F}_{0}+\rho\mathbf{I}_{n}) 
	\end{cases} \nonumber
	\end{align}
and, thus, $\mathbf{\tilde{F}}_{\rho}( d_{m} \mathbf{I}_{n} )\succeq \mathbf{\tilde{F}}_{\rho}( \mathbf{M} ) \succeq \mathbf{\tilde{F}}_{\rho}( d_{M}  \mathbf{I}_{n} )$. The provided implications are valid for $\mathbf{M}=\mathbf{D}$. In this case, one has $\mathbf{\tilde{F}}_{\rho}( \mathbf{D} ) = \mathbf{F}_{\rho}$, hence the eigenvalue bounds in \eqref{eq:upper_lower_bound_lambda_i_F_rho} can be derived for any $\rho >0$ leveraging Lem.~\ref{lem:bounds_on_eigs_F_rho}. \\
Finally, when $\cal G$ is a regular graph, it trivially holds that $d_{m}=d_{M}$ implying $\underline{\lambda}^{\mathbf{F}_{\rho}}_{i} = \lambda^{\mathbf{F}_{\rho}}_{i} = \overline{\lambda}^{\mathbf{F}_{\rho}}_{i} $,  $i \in \{1, ..., n-1\}$. \hfill {$\square$}
\end{pf}

\subsection{Convergence Properties of Scheme $\Sigma_{\rho}$}

Exploiting the spectral properties of $\mathbf{F}_{\rho}$, the rest of the section is devoted to the analysis of the role of parameter $\rho$ in the  performance of scheme $\Sigma_\rho$, by developing through three steps: first (in Prop.~\ref{prop:convergencetocentr}), the solution of $\Sigma_\rho$ is shown to converge to the centralized one; secondly (in Prop.~\ref{prop:rho_opt}), it will be discussed when to adopt either the $\Sigma_\rho$ or the $\Sigma_0$ scheme and how to tune the $\rho$ parameter; finally (in Prop.~\ref{prop:convergence_rate}), convergence rate bounds for the two schemes are obtained.

\begin{prop}\label{prop:convergencetocentr}
For any $\rho>0$, the estimate $\hat{\mathbf{x}} = \lim_{k \rightarrow +\infty}\mathbf{x}(k)  $ provided by scheme $\Sigma_{\rho}$ converges in terms of relative differences to the centralized solution $\mathbf{x}^{\star}$ in \eqref{eq:centrSol}, namely $\hat{x}_{i}-\hat{x}_{j} = x_{i}^{\star}-x_{j}^{\star}$ for all $i,j \in  \{1, \ldots, n\}$. 
\end{prop}
\begin{pf}
Since at the equilibrium it holds that 
	\begin{align}
	\hat{\mathbf{x}}  &= \mathbf{F}_{\rho} \hat{\mathbf{x}} + \mathbf{u}_{\rho} \nonumber\\
	\left( 2\mathbf{D} + \rho \mathbf{I}_{n}  \right) \hat{\mathbf{x}} &=  \left( 2\mathbf{A} + \rho \mathbf{I}_{n}  \right) \hat{\mathbf{x}} +  \tilde{\mathbf{x}} \nonumber\\
	2\mathbf{L} \hat{\mathbf{x}} &= \tilde{\mathbf{x}}, \nonumber 
	\end{align}
the solution provided by~\eqref{eq:distrUpdateVector_rho} converges to the centralized solution \eqref{eq:centrSol} in terms of relative differences, that is, up to a vector $\mathbf{y} \in \ker(\mathbf{L})$ such that $\hat{\mathbf{x}} =  \mathbf{x}^{\star} +\mathbf{y} $.\hfill$\square$
\end{pf}

Given this premise, in the following it is shown how the parameter $\rho$ can be tuned  
in order to optimize the convergence performance of the scheme $\Sigma_{\rho}$ through the minimization of the convergence rate $\mathfrak{r}_{\rho}\in   [0,1]$ of $\Sigma_\rho$, which is related to the second largest eigenvalue (in modulus) of $\mathbf{F}_{\rho}$. In particular, because of the monotonicity of $\Lambda(\mathbf{F}_\rho)$ for a given $\rho$, the convergence rate corresponds to
\begin{align}
\label{eq:convergence_rate}
\mathfrak{r}_{\rho}  = \underset{i\in \{1, \ldots, n-1\}}{\max}\vert\lambda_{i}^{\mathbf{F}_{\rho}}\vert  = {\max} \left( \vert \lambda_{1}^{\mathbf{F}_{\rho}} \vert, \vert\lambda_{n-1}^{\mathbf{F}_{\rho}} \vert \right).
\end{align}

To minimize $\mathfrak{r}_{\rho}$ in~\eqref{eq:convergence_rate}, a good and viable strategy is thus to select the parameter $\rho$ as
\begin{equation}\label{eq:min_prob_rho_star}
\rho^\star = \underset{\rho \geq 0}{\arg \min} \left\lbrace {\max} \left( \vert \lambda_{1}^{\mathbf{F}_{\rho}} \vert, \vert\lambda_{n-1}^{\mathbf{F}_{\rho}} \vert \right)  \right\rbrace
\end{equation}
The next proposition focuses on the solution of the optimization problem~\eqref{eq:min_prob_rho_star}. Note that $\rho^\star$ might be zero: this case corresponds the adoption of scheme~\eqref{eq:distrUpdateVector1}.

\begin{prop}\label{prop:rho_opt}
The solution of problem~\eqref{eq:min_prob_rho_star} exists and is unique. Defining the function $\varsigma_\mathbf M$ that acts on matrix $\mathbf M$ with ordered spectrum such that $\varsigma_\mathbf M: \mathbb{R}^{n \times n} \rightarrow \mathbb{R}$,  $\mathbf M \mapsto (\lambda^{\mathbf M }_{1}+\lambda^{\mathbf M}_{n-1})/2$, it holds that
	\begin{align}\label{eq:rhoast}
	\rho^\star = \begin{cases}
	\rho^+, \quad & \varsigma_\mathbfcal L > 1\\
	0, \quad &\varsigma_\mathbfcal L \leq 1
	\end{cases}
	\end{align}
	with
	\begin{equation}\label{eq:rho+}
	\underline{\rho}^+ = 2 (\varsigma_\mathbfcal L -1) d_{m} \leq \rho^+ \leq 2 (\varsigma_\mathbfcal L -1) d_{M} = \overline{\rho}^+.
	\end{equation}
	Furthermore, both bounds $(\underline{\rho}^+,\overline{\rho}^+)$ in \eqref{eq:rho+} coincide with $\rho^{+}$ if and only if $\mathcal{G}$ is regular.
\end{prop}
\begin{pf}
	Similarly to relation \eqref{eq:correspondance_between_lapl_and_F_eigs_}, for $\rho \geq 0$ it holds that 
	\begin{equation}\label{eq:relation_betw_muLrho_and_muFrho}
	\dfrac{1}{2}\left(\left(1-\lambda_{n-1}^{\mathbf{F}_{\rho}} \right)+\left(1-\lambda_{1}^{\mathbf{F}_{\rho}}\right)\right) =\varsigma_{\mathbfcal{L}_{\rho}} = 1-\varsigma_{\mathbf{F}_{\rho}}.
	\end{equation}
Thanks to Lem.~\ref{lem:bounds_on_eigs_F_rho}, $\varsigma_{\mathbf{F}_{\rho}} $ is proven to be a continuous strictly increasing function of the parameter $\rho$, thus ensuring the existence of a unique solution for~\eqref{eq:min_prob_rho_star}. This also implies that $\varsigma_{\mathbf{F}_{\rho}} > \varsigma_{\mathbf{F}_{0}}$ for $\rho>0$.  
	\\
Then, if $\varsigma_{\mathbf{F}_0} \geq 0 $, namely $\varsigma_\mathbfcal L \leq 1$, it follows that $\varsigma_{\mathbf{F}_{\rho}} > 0$. 
Therefore, due to~\eqref{eq:relation_betw_muLrho_and_muFrho}, the convergence rate results to be $\mathfrak{r}_{\rho} = \vert \lambda_{1}^{\mathbf{F}_{\rho}}\vert$ and it is trivially minimized for $\rho = 0$. Conversely, if $\varsigma_{\mathbf{F}_{0}} < 0 $, namely $\varsigma_\mathbfcal L > 1$, the condition $\varsigma_{\mathbf{F}_{\rho}} = 0$, corresponding to the optimal scenario wherein $\vert \lambda_{1}^{\mathbf{F}_{\rho}} \vert= \vert\lambda_{n-1}^{\mathbf{F}_{\rho}} \vert$ may occur. In particular, for Bolzano's theorem, it follows that the condition $\varsigma_{\mathbf{F}_{\rho}} = 0$ is fulfilled only for one value of $\rho$ ($\rho^+$). \\ 
Finally, resting upon Prop.~\ref{prop:bounds_on_eigs_F_rho}, inequalities~\eqref{eq:rho+} can be derived by imposing 
	\begin{equation*}
	(\underline{\lambda}^{\mathbf{F}_{\rho^{\star}}}_{1}+\underline{\lambda}^{\mathbf{F}_{\rho^{\star}}}_{n-1})/2 \leq \varsigma_{\mathbf{F}_{\rho^{\star}}} \leq (\overline{\lambda}^{\mathbf{F}_{\rho^{\star}}}_{1}+\overline{\lambda}^{\mathbf{F}_{\rho^{\star}}}_{n-1})/2
	\end{equation*}
	and exploiting the fact that $\varsigma_{\mathbf{F}_{\rho\star}} = 0$ only when $\varsigma_{\mathbf{F}_{0}} < 0 $. Trivially, the bounds~\eqref{eq:rho+} are met  for regular graphs since $d_m = d_M$ for these topologies. \hfill$\square$
\end{pf}

\begin{Rem} \label{Remark1} Prop.~\ref{prop:rho_opt} suggests when it is useful to adopt the PP-based solution~\eqref{eq:distrUpdateVector_rho} basing on the graph describing the given multi-agent system through $\varsigma_\mathcal{L}$. By definition, this quantity depends on the network connectivity and similarity to bipartite graph. Hence, when $\varsigma_\mathbfcal L > 1$, the corresponding topology tends to circulate the information between node groups. In this case, the choice $\rho^\star = \rho^+$ allows to improve the information management. On the other hand,  when $\varsigma_\mathbfcal L \leq 1$, either $\rho^\star$ is set as an arbitrary relatively small value or  the scheme $\Sigma_{0}$ has to be preferred, since the addition of self loop in the graph does not improve (and often worsen) the estimation speed.
\end{Rem} 

To conclude, the next proposition discusses the bounds of the convergence rate for the optimal selection of $\rho$ in~\eqref{eq:rhoast}

\begin{prop}\label{prop:convergence_rate}
	If $\varsigma_{\mathbfcal L} \leq 1 $, the rate of convergence $\mathfrak{r}_{\rho^{\star}}$ for scheme $\Sigma_{\rho^{\star}} = \Sigma_{0}$ is yielded by $\mathfrak{r}_{\rho^{\star}} =  1-\lambda_{1}^{\mathbfcal L} \in (0,1) $. Otherwise, if $\varsigma_{\mathbfcal L} > 1 $, it is bounded as follows
	\begin{align}
	\mathfrak{r}_{\rho^{\star}} & \leq \max\left( - \underline{\lambda}_{n-1}^{\mathbf{F}_{\underline{\rho}^{+}}} , \overline{\lambda}_{1}^{\mathbf{F}_{\overline{\rho}^{+}}}\right) <1, \nonumber\\
	\mathfrak{r}_{\rho^{\star}} & \geq \max\left( -\min\left(0, \overline{\lambda}_{n-1}^{\mathbf{F}_{\overline{\rho}^{+}}} \right) , \max\left(0 , \underline{\lambda}_{1}^{\mathbf{F}_{\underline{\rho}^{+}}} \right) \right) \!\geq \!0. \nonumber
	\end{align}
In particular, $\mathfrak{r}_{\rho^{\star}} = 0$ if and only if $\cal G$ is the complete graph.
\end{prop}
\begin{pf}
	The result follows immediately from Prop. \ref{prop:bounds_on_eigs_F_rho}, Prop. \ref{prop:convergencetocentr}, Prop. \ref{prop:rho_opt}, the fact that if $\lambda_{n-1}^{\mathbf{F}_{0}}=-1$ then $\lambda_{n-1}^{\mathbf{F}_{\rho}} \geq  (\rho - 2d_{M})/(\rho+2d_{M})  > -1$ for $\rho>0$, and the general fact that $0 < \lambda_{1}^{\mathbfcal L} \leq 1$ holds for a non-complete connected graph $\cal G$. 
	Moreover, since nontrivial eigenvalues of $\mathbfcal L$ are all equal to $n(n-1)^{-1}$ if and only if $\cal G$ is the complete graph, the assignment of $\rho^{+}=2$ to the parameter $\rho$ allows to place all the eigenvalues of $\mathbf{F}_{\rho}$ different from $1$ in $0$, achieving superlinear convergence ($\mathfrak{r}_{\rho}^{\star} = 0$). \hfill$\square$
\end{pf}

\section{Numerical Results}\label{sec:numerical_simulation}
In this section, a comparison between $\Sigma_\rho$ and $\Sigma_0$ is provided accounting for different network topologies and focusing the attention on the convergence properties of the two decentralized schemes. 

In each test, the number of iterations is fixed to $\bar{k} = 20$ and the two following  performance indexes are considered: 
\begin{align}\label{eq:effective_conv_rate}
\mathfrak{r}_{e} &= \dfrac{1}{\bar{k}-2k_{neg}}\sum\limits_{k=k_{neg}}^{\bar{k}-k_{neg}-1} \dfrac{\left\| \mathbf{x}(k+1)-\mathbf{x}^{\star}\right\|_{2}}{\left\| \mathbf{x}(k)-\mathbf{x}^{\star}\right\|_{2}}, \\
\label{eq:MSE}
\mathrm{MSE} &= \dfrac{1}{n}\sum\limits_{i=1}^{n}\left(x_{i}(\bar{k})-x_{i}^{\star}\right)^{2}.
\end{align}
The former evaluates the effective convergence rate neglecting the first and last $k_{neg} = 2$ iterations and taking into account the average estimation error ratio, while the latter corresponds to the final mean square error\footnote{In \eqref{eq:effective_conv_rate}-\eqref{eq:MSE}, the state components are normalized in difference by, for instance, subtracting the first component to all the others.}.

\subsection{Case study: nonregular bipartite topology}

\begin{figure}[t!]
	\centering
	\subfigure[]{\includegraphics[width=0.35\columnwidth]{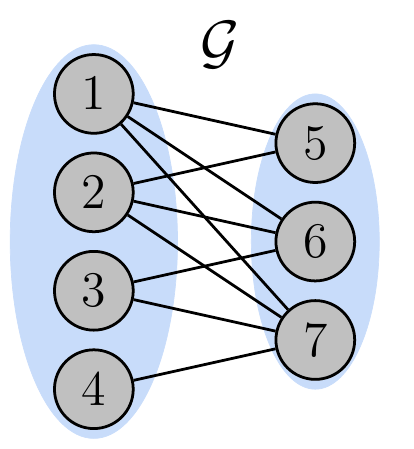}\label{fig:graph}}
	\subfigure[]{\includegraphics[width=0.65\columnwidth]{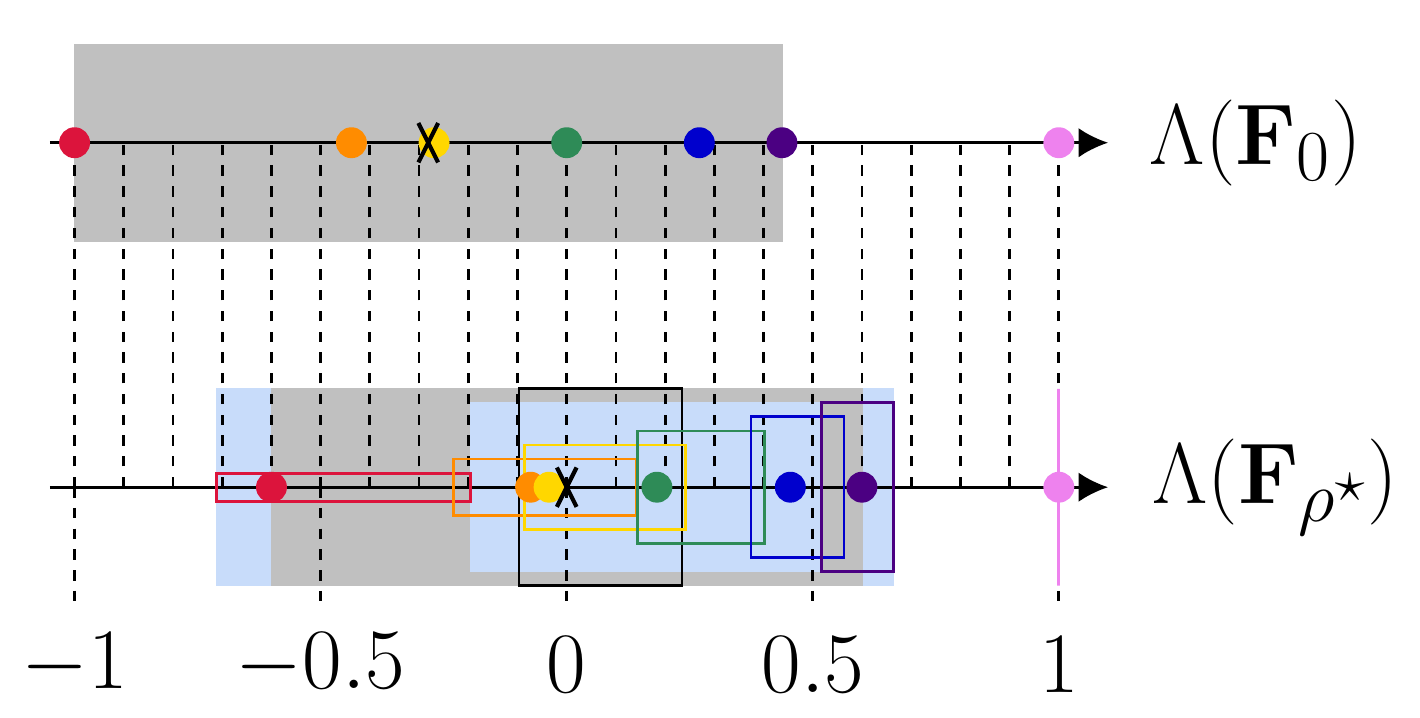}\label{fig:eigen}}
	\caption{(a) Nonregular bipartite graph with $7$ nodes. (b) Spectral mapping due to penalty parameter tuning: 
the optimization of $\rho$ allows the zeroing of the spectral centroid (black cross) and the reduction of the nonperipheral spectral radius (gray rectangle).} 
	\label{fig:sim0}		
\end{figure}

In this first test, the bipartite network in Fig. \ref{fig:graph} is considered, assuming that the optimal solutions difference $x^{\star}_{i+1}-x^{\star}_{i}$ is equal for $i \in \{1, \ldots, n-1\}$ and that the relative measurements are corrupted by additive uniformly distributed noise.

Fig.~\ref{fig:eigen} shows how the eigenvalues of the matrix $\mathbf{F}_{0}$ are mapped into the spectrum of $\mathbf{F}_{\rho}$, when the penalty parameter $\rho$ corresponds to the solution $ \rho^{\star}$ of~\eqref{eq:min_prob_rho_star}. Note that, according to Lem.~\ref{lem:bounds_on_eigs_F_rho} and Prop. \ref{prop:bounds_on_eigs_F_rho}, all the the eigenvalues $\lambda_{i}^{\mathbf{F}_{\rho^{\star}}}$ are located within a certain neighborhood and maintain the same order w.r.t. the elements in $\Lambda(\mathbf{F}_{0})$. In addition, since it is possible to compute the spectrum of the normalized Laplacian $\Lambda(\mathbfcal{L})$, Prop. \ref{prop:rho_opt} entails that $\rho^{\star}$ belongs to $[\underline{\rho}^{\star},\overline{\rho}^{\star}]\simeq[0.5626,2.2506]$,  hence, the convergence rate $\mathfrak{r}_{\rho^{\star}}$ results in the interval $[\underline{\mathfrak{r}}_{\rho^{\star}},\overline{\mathfrak{r}}_{\rho^{\star}}]\simeq[0.5184,0.7118]$.

\begin{figure}[t!]
	\centering
	\includegraphics[trim={0 1.25cm 0 0},clip,width=1\columnwidth]{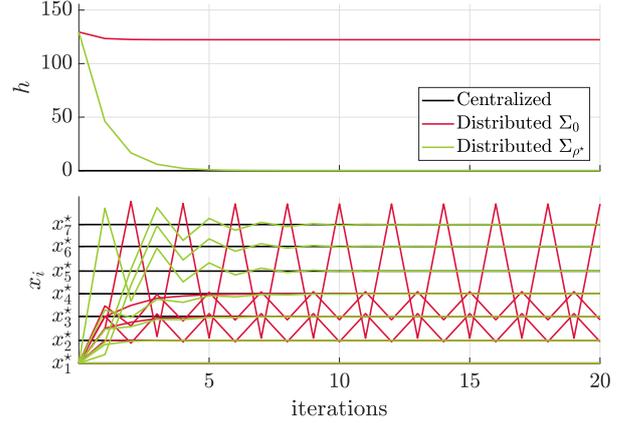}
	\caption{(top) Cost function trend: the convergence is not attained when the scheme $\Sigma_{0}$ is adopted since $\mathcal{G}$ is bipartite. (bottom) Iterative schemes behavior: the  state estimates  oscillate around their optimal values for $\Sigma_0$,  while  they converge to the optimum for $\Sigma_{\rho^{\star}}$.} 
	\label{fig:sim1}		
\end{figure}

From Fig.~\ref{fig:sim1}, one can note that the convergence is not ensured adopting the scheme $\Sigma_{0}$: the convergence rate trivially results $\mathfrak{r}_{0} = 1$.
In detail, neither the cost function (top panel) \textquotedblleft approaches" zero nor the estimations (bottom panel) converge towards the optimal solutions. This issue is easily overcome when the PP-based scheme is implemented: the choice of the penalty $\rho^{\star} = 1.3469$ not only guarantees convergence but also speeds up the estimation dynamics ($\mathfrak{r}_{\rho^{\star}} = 0.5999 < \mathfrak{r}_{0}$). 
In addition, accounting for the performance indexes~\eqref{eq:effective_conv_rate}-\eqref{eq:MSE}, the scheme $\Sigma_{\rho^{\star}}$ yields $\mathfrak{r}_{e} = 0.0471$ and $\mathrm{MSE} = 2.7485\cdot 10^{-4}$, whereas, for the scheme $\Sigma_{0}$ only the final mean square error can be evaluate since convergence is not attain and this results considerably higher as compared to $\Sigma_{\rho^\star}$ being $\mathrm{MSE} = 5.8291$.

\begin{table*}[t!]
	\caption{Summary of the validation test for the proposed iterative schemes in a heterogeneous topology framework.}
	\label{tab:diff_topologies}
	\begin{center}
	\begin{tabular}{|c|c|c|c|c|c|c|c|c|}
		\cline{1-9}
		\multicolumn{1}{|c||}{Topology} & \cellcolor{blue!25}$K_{36}$ & $C_{36}(1,2)$ & $\mathcal{R}_{36}$ & \cellcolor{gray!25}$\Gamma_{36}(3)$ & \cellcolor{green!25}$SW_{9,27}$ & \cellcolor{gray!25}$\mathcal{S}_{36}$ & $\mathcal{B}^{+}_{4}$ & $\mathcal{B}^{+}_{6}$ \\ \hline\hline
		\multicolumn{1}{|c||}{Regular} & yes & yes & yes & yes & no & no & no & no \\ \hline
		\multicolumn{1}{|c||}{Bipartite} & no & no & no & \cellcolor{gray!25} yes & no & \cellcolor{gray!25} yes & no & no \\ \hline \hline
		\multicolumn{1}{|c||}{$\mathrm{dens}$} & $1$ & $0.1143$ & $0.0857$ & $0.0857$ & $0.6159$ & $0.0556$ & $0.0667$ & $0.0159$ \\ \hline
		\multicolumn{1}{|c||}{$\phi$} & $1$ & $9$ & $6$ & $6$ & $3$ & $2$ & $8$ & $12$ \\ \hline
		\multicolumn{1}{|c||}{$d_{av}$} & $35$ & $4$  & $3$ & $3$ & $21.5556$ & $1.9444$ & $2$ & $2$ \\ \hline
		\multicolumn{1}{|c||}{$d_{m}$} & $35$ & $4$ & $3$ & $3$ & $8$ & $1$ & $1$ & $1$ \\ \hline
		\multicolumn{1}{|c||}{$d_{M}$} & $35$ & $4$ & $3$ & $3$ & $27$ & $35$ & $3$ & $3$ \\ \hline
		\multicolumn{1}{|c||}{$\lambda_{1}^{\mathbfcal{L}}$} & $1.0286$ & $0.0377$ & $0.1548$ & $0.1181$ & $0.0133$ & $1$ & $0.0261$ & $0.0050$ \\ \hline
		\multicolumn{1}{|c||}{$\lambda_{n-1}^{\mathbfcal{L}}$} & $1.0286$ & $1.5567$ & $1.9407$ & $2$ & $1.1456$ & $2$ & $1.9888$ & $1.9980$ \\ \hline
		\multicolumn{1}{|c||}{$\varsigma_{\mathbfcal L}$} & \cellcolor{blue!25}$1.0286$ & $0.7972$ & $0.9720$ & \cellcolor{gray!25}$1.0590$ & \cellcolor{green!25}$0.5795$ & \cellcolor{gray!25}$1.5$ & $1.0074$ & $1.0015$ \\ \hline
		\multicolumn{1}{|c||}{$\underline{\rho}^{\star}$} & $2$ & $0$ & $0$ & $0.3542$ & $0$ & $1.0000$ & $0.0148$ & $0.0030$ \\ \hline
		\multicolumn{1}{|c||}{$\rho^{\star}$} & \cellcolor{blue!25}$2$ & $0$ & $0$ & \cellcolor{gray!25}$0.3542$ & \cellcolor{green!25}$0$ & \cellcolor{gray!25}$1.8972$ & $0.0283$ & $0.0059$ \\ \hline
		\multicolumn{1}{|c||}{$\overline{\rho}^{\star}$} & $2$ & $0$ & $0$ & $0.3542$ & $0$ & $35.0000$ & $0.0445$ & $0.0090$ \\ \hline
		\multicolumn{1}{|c||}{$\underline{\mathfrak{r}}_{\rho^{\star}}$} & $0$ & $0.9623$ & $0.9131$ & $0.8885$ & $0.9867$ & $0.0264$ & $0.9740$ & $0.9950$\\ \hline
		\multicolumn{1}{|c||}{$\mathfrak{r}_{\rho^{\star}}$} & $0$ & $0.9623$ & $0.9131$ & $0.8885$ & $0.9867$ & $0.4868$ & $0.9741$ & $0.9950$\\ \hline
		\multicolumn{1}{|c||}{$\overline{\mathfrak{r}}_{\rho^{\star}}$} & $0$ & $0.9623$ & $0.9131$ & $0.8885$ & $0.9867$ & $0.9472$ & $0.9794$ & $0.9961$ \\ \hline
		\multicolumn{1}{|c||}{$\mathfrak{r}_{0}$} & $0.0286$ & $0.9623$ & $0.9131$ & $1$ & $0.9867$ & $1$ & $0.9888$ & $0.9980$ \\ \hline \hline
		\multicolumn{1}{|c||}{$\mathfrak{r}_{e}(\Sigma_{0})$} & $5.2521\cdot 10^{-5}$ & $0.2648$ & $0.1940$ & $-$ & \cellcolor{green!25}$0.7344$ & $-$ & $0.3149$ & $0.4343$ \\ \hline
		\multicolumn{1}{|c||}{$\mathfrak{r}_{e}(\Sigma_{\rho^{\star}})$} & \cellcolor{blue!25}$7.6734\cdot 10^{-16}$ & $0.2648$ & $0.1940$ & \cellcolor{gray!25}$0.0399$ & $0.7344$ & \cellcolor{gray!25}$0.0289$ & $0.2863$ & $0.4336$ \\ \hline \hline
		\multicolumn{1}{|c||}{$\mathrm{MSE}(\Sigma_{0})$} & $1.6732\cdot 10^{-14}$ & $3.9255$ & $0.8262$ & $0.2566$ & $14.331$ & $18.5849$ & $2.2622$ & $12.1184$ \\ \hline
		\multicolumn{1}{|c||}{$\mathrm{MSE}(\Sigma_{\rho^{\star}})$} & $1.6711\cdot 10^{-14}$ & $3.9255$ & $0.8262$ & $0.0296$ & $14.331$ & $1.1920\cdot 10^{-5}$ & $1.9135$ & $12.0192$ \\ \hline
	\end{tabular}
	\end{center}
\end{table*}

\subsection{Other Topologies}

In the following, the comparison between $\Sigma_0$ and $\Sigma_\rho$ is carried out accounting for many different network topologies. 
In detail, the performances are evaluated for the networks considered in~\citep{fabris2019distributed}, namely the complete graph $K_{36}$, the circulant graph $C_{36}(1,2)$, the Ramanujan graph $\mathcal{R}_{36}$ and the Cayley graph $\Gamma_{36}(3)$, all with $n=36$ nodes. Furthermore, the following networks are also investigated: the star graph $\mathcal{S}_{36}$ with $n=36$, i.e., a tree with $35$ nodes characterized by a degree of $1$ linked to a common root with $35$ incidences;
the small-world $SW_{9,27}$ with $n=36$, constituted by two complete graphs $K_{9}$ and $K_{27}$ linked together by one single edge;
the complete binary trees $\mathcal{B}^{+}_{4}$ and $\mathcal{B}^{+}_{6}$ having $n=32$ and $n=128$ nodes, respectively, and characterized by an extra connection between the root and one of the leaves.

It is worth to recall that some of these basic topologies set the foundations for the development of key applications as camera networks, smart grids and social networks \citep{chen2019control}. As a result, Tab.~\ref{tab:diff_topologies} 
summarizes the parameters and all the main properties related to the convergence behaviors, jointly with some other graph-related quantities, as the \textit{average degree} $d_{av}(\mathcal{G}) = n^{-1} \sum_{i=1}^{n}d_{i}$, the \textit{density} $\mathrm{dens}(\mathcal{G}) = 2\vert\mathcal{E}\vert/(n(n-1))$ and the \textit{diameter} $\phi(\mathcal{G}) = \max{\left\lbrace\min(\vert l_{ij}\vert-1) \right\rbrace}$, where $\vert l_{ij}\vert$ is the number of vertices belonging to the path that links nodes $i$ and $j$.

From the data reported in this table, some considerations are in order.
\begin{itemize}
	\item $K_{36}$: the estimation performance is good for both $\Sigma_{0}$ and $\Sigma_{\rho^{\star}}$, although the latter method allows for a faster convergence;
	\item $C_{36}(1,2)$ and $\mathcal{R}_{36}$: for these specific (regular) graphs the performances of $\Sigma_{0}$ and $\Sigma_{\rho^{\star}}$ are the same, for both indexes $\mathfrak{r}_{e}$ and $\mathrm{MSE}$ (as a further remark, note that for some circulant graphs, as the ring graphs, the performance results better by employing the PP-based approach);
	\item $\Gamma_{36}(3)$ and $\mathcal{S}_{36}$: for the bipartite nature of these graphs, only the PP-based estimation can be applied successfully ($\Gamma_{36}(3)$ is regular and $\mathcal{S}_{36}$ is not);
	\item $SW_{9,27}$: the two update rules $\Sigma_{0}$ and $\Sigma_{\rho^{\star}}$ perform equally, for this nonregular graph;
	\item $\mathcal{B}^{+}_{4}$ and $\mathcal{B}^{+}_{6}$: estimation performances are similar between the two methods, nonetheless, the PP-based procedure obtains better results both in estimation speed and estimation accuracy.
\end{itemize}

By recalling Remark~\ref{Remark1}, colored cells in Tab. \ref{tab:diff_topologies} highlight how bipartite and complete graphs improve information spreading due to their connectivity patterns, (with $\varsigma_{\mathbfcal L}>1$, using $\Sigma_{\rho^{\star}}$)
while small-world networks tend to isolate information in node clusters (with $\varsigma_{\mathbfcal L}<1$, using $\Sigma_0$).



\section{Conclusions}\label{sec:conclusions}

This work focuses on system state estimation problem consisting in the computation of the optimal state estimation for all the components of a networked multi-agent system characterized by a set of noisy relative measurements. 
To solve this problem, a distributed iterative PP-based scheme is provided in the form of linear state-space model. The proposed solution is proven to converge in terms of relative differences to the corresponding centralized solution. 
In particular, its convergence features rests upon the spectral properties of stochastic matrices strictly related to the network topology and the normalized Laplacian matrix of a graph. Lower and upper bounds for the convergence rate of the provided PP scheme are reported along with the indications to compute an optimal value for the penalty parameter. Numerical simulations run in different scenarios validate the proposed estimation scheme and support the theoretical considerations on its performance.\\
Future directions for this work are represented by a performance comparison between the PP-based scheme, the spectral-based scheme $\Sigma_{\eta}$ derived in \citep{fabris2019distributed} and other schemes, as well as their generalization. 



\bibliography{biblio_v2}           
\end{document}